\documentclass[11pt]{article}
\usepackage{moriond}

\newcommand{\ome}[2][]{\ensuremath{
    \Omega_{#1}^{#2}}}
\newcommand{\weff}{\ensuremath{\overline{w}_0}}
\newcommand{\omebar}[2][]{
    \overline{\Omega}_{\rm #1}^{#2}}

\def\be{\begin{equation}}
\def\ee{\end{equation}}
\def\bea{\begin{eqnarray}}
\def\eea{\end{eqnarray}}

\begin{document}
\vspace*{4cm}
\title{The Influence of Quintessence on the Separation of CMB Peaks}

\author{Michael Doran, Matthew Lilley, Jan Schwindt and Christof Wetterich}

\address{Institut f\"ur Theoretische Physik der Universit\"at Heidelberg\\ 
    Philosophenweg 16, D-69120 Heidelberg, Germany}
  
  \maketitle\abstracts{A hypothetical dark energy component may have
    an equation of state that is different from a cosmological
    constant and possibly even changing in time. The spacing of the
    cosmic microwave background peaks is sensitive to the ratio of the
    horizon sizes today and at last scattering and therefore to the
    details of the cosmological evolution.  Together with independent
    measurements of today's cosmological parameters, this restricts
    quintessence models in the epoch before and during decoupling.
    For fixed $\Omega^{\tiny\textrm{b}}_0$, $\Omega^{\tiny \textrm{cdm}}_0$ and $h$, we give an
    analytic estimate of the spacing that depends on $\Omega^{\phi}_{\tiny\rm ls}$,
    $\Omega^{\phi}_0$ and an effective equation of state of the dark
    energy component.
  }

\section{Introduction}
Quintessence \cite{orig,peebl} is a form of homogeneous energy associated with a scalar field 
having  a time varying equation of state. 
If quintessence  played a role in the epoch of last scattering, it may have 
left imprints in the Cosmic Microwave Background (CMB) fluctuations. 
The position of the $m$-th peak in the angular momentum spectrum of the
CMB can be parameterized \cite{white} by
\[
l_m = \Delta l (m - \phi_m), 
\]
where $\Delta l$ is the average spacing between the peaks and the
shift $\phi_m$ is a small quantity with absolute value less than
$0.4$, typically.  For a flat universe and adiabatic initial conditions, 
$\Delta l$ is given 
by the simple formula \cite{hu95,hu97}
\begin{equation}
\label{platz}
\Delta l = \pi  \frac{\tau_0 - \tau_{\rm ls}}{s} = \pi \frac{\tau_0 -
  \tau_{\rm ls}}{\bar c_s \tau_{\rm ls}}.
\end{equation}
Here $\tau_0$ and $\tau_{\rm ls}$ are the conformal time today and at
last scattering (which are equal to the particle horizons) and
$\tau=\int {\rm d}t \ a^{-1}(t)$, with cosmological scale factor $a$.
The sound horizon at last scattering $s$ is related to $\tau_{\rm ls}$
by $s =\bar c_s\tau_{\rm ls}$, where the average sound speed before
last scattering is approximately $\sqrt{1/3}$.

In this paper we will show that the spacing of the CMB peaks can be
sensitive to quintessence.
\begin{figure}[!ht]
\begin{center}
\input{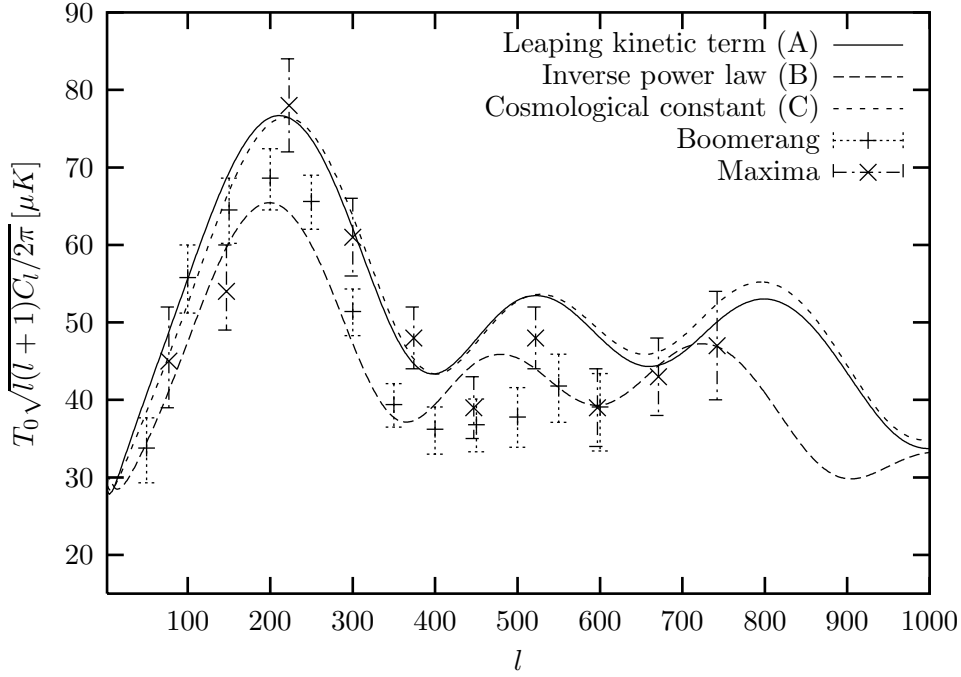}
\end{center}
\caption{The CMB Spectrum for $\Lambda$-CDM (model C), leaping kinetic term
  (model A) $\!^6$ and inverse power law (model B) quintessence universes
with \ome[0]{\phi}=0.6.  The data points from the Boomerang $\!^7$
and Maxima $\!^8$\vspace{-0.2em} experiments are shown for
  reference.}
\label{cmb}
\end{figure}

\section{Analytic estimate for $\Delta l$}
We present here a quantitative discussion of the mechanisms which
determine the spreading of the peaks.  A simple analytic formula
permits us to relate $\Delta l$ directly to three characteristic
quantities for the history of quintessence, namely the fraction of
dark energy today, $\ome[0]{\phi}$, the averaged ratio between dark
pressure and dark energy, $\weff = \langle p^{\phi} / \rho^{\phi}\rangle_0$, 
and the averaged quintessence fraction before
last scattering, $\omebar[ls]{\phi}$ (for details of the averaging
see Appendix).  We compare our estimate with an explicit numerical
solution of the relevant cosmological equations using CMB-FAST
\cite{se96}.  For a given model of quintessence the computation of the
relevant parameters $\ome[0]{\phi}, \weff$ and $\omebar[ls]{\phi}$
requires the solution of the background equations.  Our main
conclusion is that future high-precision measurements of the location
of the CMB-peaks can discriminate between different models of dark
energy if some of the cosmological parameters are fixed by independent
observations (see fig. \ref{cmb}).  It should be noted here that a likelihood analysis of
the kind performed in \cite{bond00}, where $w$ is assumed to be
constant throughout the history of the Universe, would not be able to
extract this information as it does not allow $\omebar[ls]{\phi}$ to
vary.  We point out that for time-varying $w$ there is no direct
connection between the parameters $\weff$ and $\omebar[ls]{\phi}$,
i.e. a substantial $\omebar[ls]{\phi}$ (say 0.1) can coexist with
rather large negative $\weff$.  We perform therefore a three parameter
analysis of quintessence models and our work goes beyond the
investigation for constant $w$ in \cite{steinhardt}.

The average spacing is calculated \cite{mat} by integrating the Friedmann equation,
assuming a constant fraction $\omebar[ls]{\phi}$ of quintessence energy
before last scattering and the effective equation of state $\weff$ for later times:

\begin{equation}
\label{sep}
\Delta l =   \pi \bar c_s^{-1} \left[
  \frac{F(\ome[0]{\phi},\weff)}{\sqrt{1-{\omebar[ls]{\phi}}}}
      \left \{ \sqrt{a_{\rm ls} + {\ome[0]{\rm r}\over 1 
- \ome[0]{\phi}}} - \sqrt{{\ome[0]{\rm r}\over 1 - \ome[0]{\phi}}} \right \} ^{-1} - 1 \right ],
\end{equation}
with
\begin{equation} \label{F_int}
F(\ome[0]{\phi},\weff) = \frac{1}{2} \int_0^1 \textrm{d}a \left( a +
  \frac{\ome[0]{\phi}}{1-\ome[0]{\phi}} \, a ^{(1 - 3 \weff)} +
  \frac{\ome[0]{\rm r}(1-a)}{1-\ome[0]{\phi}} \right)^{-1/2},
\end{equation}
and today's radiation component $\ome[0]{\rm r} h^2= 4.2\times 10^{-5},
a_{\rm ls}^{-1} \approx 1100$ and $\overline{c}_s = 0.52$. 

An alternative to the spacing between the peaks is the ratio of any
two peak (or indeed trough) locations.  After last scattering the CMB
anisotropies simply scale \footnote{Except for the ISW effect, which is 
slowly varying with $l$ and rather small.}
according to the geometry of the Universe -- taking the ratio of two
peak locations factors out this scaling and leaves a quantity which is
sensitive only to pre-last-scattering physics.  As can be seen in
\mbox{Table 1}, (spatially-flat) models with negligible $\omebar[ls]{\phi}$
all have $l_2 / l_1 \approx 2.41$ for the parameters given in Table
\ref{lollie}.  The dependence of this ratio on the other cosmological
parameters can be computed numerically \cite{doran}, and thus a
deviation from the predicted value could be a hint of time-varying
quintessence.

\begin{table}[!ht]
\begin{center}
\begin{tabular}{cccccccc}
\hline
$\bf {\omebar[ls]{\phi}}$ & $\weff$ & $\bf l_1$ & $\bf l_2$ & $\bf
l_2/l_1$ & $\bf  \Delta l ^{estim.}$ & $\bf \Delta l^{num.}$ & $\bf
\sigma_8$ \\ 
\hline
\multicolumn{8}{c}{Leaping kinetic term
  (A), \ome[0]{\phi} = 0.6} \\
\hline
$8.4 \times 10^{-3} $ &  $-0.76$ &$215$ & $518$ & $ 2.41$ & $292$ &
$291$ &$0.86$\\ 
$0.03$ & $-0.69$& $214$ & $520$ & $2.43$ & $294$ & $293$ &$0.78$\\
$0.13$ & $-0.45$&  $211$ & $523$ & $2.48$ & $299$  & $300$&$0.47$ \\
$0.22$ & $-0.32$ & $207$ & $524$ & $2.53$ & $302$ & $307$ &$0.29$\\
\hline
\multicolumn{8}{c}{Inverse power law
  potential (B), \ome[0]{\phi} = 0.6} \\
\hline
$8.4 \times 10^{-8}$ & $-0.37$&$199$ & $480$ & $ 2.41$ &  $271$ &
$269$ &$0.61$\\ 
$9.9 \times 10^{-2}$ & $-0.13$&$178$ & $443$ & $2.49$ & $252$ &
$252$&$0.18$ \\ 
$0.22$ & $-8.1\times 10^{-2}$&  $172$ & $444$ & $2.58$ & $257$ &
$257$&$0.09$ \\ 
\hline
\multicolumn{8}{c}{Pure exponential
  potential, \ome[0]{\phi} = 0.6} \\
\hline
$0.70$ &  $7\times 10^{-3}$ &  $190$ & $573$ & $3.02$ & $368$ &
$377$&$0.01$ \\ 
\hline
\multicolumn{8}{c}{Pure exponential potential,
 $\ome[0]{\phi}=0.2$} \\ 
\hline
$0.22$ & $4.7\times 10^{-3}$ & $194$ & $490$ & $2.53$ & $282$ & $281$
&$0.38$\\     
\hline
\multicolumn{8}{c}{Cosmological constant
  (C), \ome[0]{\phi} = 0.6} \\
\hline
$0$ & $-1$&  $219$ & $527$ & $ 2.41$ & $296$ & $295$&$0.97$ \\
\hline
\multicolumn{8}{c}{Cold Dark Matter - no
  dark energy, \ome[0]{\phi} = 0} \\
\hline
$0$ & $ - $&  $205$ & $496$ & $ 2.42$ & $269$ & $268$ &$1.49$ \\
\hline
\end{tabular} \caption{Location of the first two CMB peaks $l_1,\ l_2$
  for several models of dark energy.  We also show the analytic (from
  Equation (\ref{sep}) and numerical (from CMB-FAST) average spacing
  of the peaks, the ratio $l_2/l_1$ of the peak locations and
  $\sigma_8$, the normalisation of the power spectrum on scales of
  $8h^{-1}$Mpc.}
\end{center}
\label{lollie}
\normalsize
\end{table}

\section{Conclusion}
Structure formation restricts \ome[]{\phi} to be less than about $0.2$
until very recently (see table \ref{lollie}). Big bang nucleosynthesis
gives about the same bound. When MAP \cite{map} has captured the CMB
spectrum up to the third peak, it should - together with other means
such as Supernovae data, clustering and lensing - be possible to
narrow the space of cosmic parameters such that bounds on quintessence
in a range of redshift $z = 10^5 .. 10^3$ and on $\weff$ can be
derived. The phase shifts $\phi_m$ and hence the peak ratios will
single out the recombination physics of quintessence, whereas $\Delta
l$ as a combination of both recombination and late time cosmology will
help in determining $\weff$.

\section*{Appendix}
The effective equation of state is defined as \ome[]{\phi} weighted 
$\tau$ average:
\begin{equation}
\label{w_eff}
\weff =  \int_0^{\tau_0} \ome[]{\phi}(\tau) w(\tau) \textrm{d} \tau 
\times \left(  \int_0^{\tau_0} \ome[]{\phi}(\tau) \textrm{d} \tau
\right)^{-1}.
\end{equation}
Similarly, $\omebar[ls]{\phi} \equiv \tau_{\rm ls}^{-1}
\int_0^{\tau_{\rm ls}} \ome[]{\phi}(\tau) \textrm{d}\tau$ is just
the $\tau$ average.

\begin{table}[!ht]
\begin{center}
\caption{Symbols, their meanings and numerical values used.}
\vspace{1ex}
\begin{tabular}{|clc|}
\hline
Symbol & Meaning & Value \\
\hline
$a(\tau)$ & scale factor, normalised to unity today &\\
$a_{\rm ls}$ & scale factor at last scattering & $1100^{-1}$ \\
$h_0$ & Hubble parameter today $H_0 = 100\,h_0\,\textrm{km
  s}^{-1} {\rm Mpc}^{-1}$ &  $0.65$ \\
\ome[0]{\rm r} & relativistic $\Omega$ today & $9.89\times 10^{-5}$ \\
\ome[0]{\rm b} & baryon $\Omega$ today & $0.05$ \\
$\bar c_s$ & $\tau$-averaged sound speed until last scattering & $ 0.52$ \\ 
$n$ & spectral index of initial perturbations & $1$ \\[0.5ex]
\hline
\end{tabular}
\label{symb}
\end{center}
\end{table}

\end{document}